\documentclass[aps,prd,reprint,superscriptaddress]{revtex4-1}
\pdfoutput=1
\usepackage{graphicx}
\usepackage{amsmath}

\begin{document}


\title{Measuring the Redshift of Reionization with a Modest Array \\ of Low-Frequency Dipoles}


\author{Jonathan M. Bittner}
\email[]{jbittner@physics.harvard.edu}
\affiliation{Jefferson Laboratory of Physics, Harvard University, 17 Oxford St, Cambridge, MA 02138}
\affiliation{Institute for Theory \& Computation, Harvard-Smithsonian CfA, 60 Garden Street, Cambridge, MA 02138}

\author{Abraham Loeb}
\email[]{aloeb@cfa.harvard.edu}
\affiliation{Institute for Theory \& Computation, Harvard-Smithsonian CfA, 60 Garden Street, Cambridge, MA 02138}

\date{\today}

\begin{abstract}
The designs of the first generation of cosmological 21-cm observatories are split between single dipole experiments which
integrate over a large patch of sky in order to find the global (spectral)
signature of reionization, and interferometers with arcminute-scale
angular resolution whose goal is to measure the 3D power spectrum of
ionized regions during reionization.  We examine whether intermediate
scale instruments with complete Fourier ($uv$) coverage are capable of
placing new constraints on reionization.  We find that even without
using a full power spectrum analysis, the global redshift of
reionization, $z_{\rm reion}$, can in principle be measured from the
variance in the 21-cm signal among multiple beams as a function of
frequency at a roughly 1 degree angular scale.  At this scale, the
beam-to-beam variance in the differential brightness temperature peaks
when the average neutral fraction was $\sim 50\%$, providing a
convenient flag of $z_{\rm reion}$. We choose a low angular resolution
of order $1^{\circ}$ to exploit the physical size of the ionized
regions and maximize the signal-to-noise ratio.  Thermal noise,
foregrounds, and instrumental effects should also be manageable at
this angular scale, as long as the $uv$ coverage is complete within
the compact core required for low-resolution imaging.  For example, we
find that $z_{\rm reion}$ can potentially be detected to within a
redshift uncertainty of $\Delta z_{\rm reion}\lesssim 1$ in $\gtrsim 500$
hours of integration on the existing MWA prototype (with only
$32\times 16$ dipoles), operating at an angular resolution of $\sim 1^{\circ}$ and a spectral resolution of 2.4 MHz. 
\end{abstract}


\maketitle


\section{Introduction \label{intro}}
The 21-cm line from neutral hydrogen is a powerful probe of the
high-redshift universe.  Its observations could potentially constrain
the properties of the first galaxies \cite{2009MNRAS.397.1454B, LoebText, 2009astro2010S..83F}, the cosmological initial conditions
\cite{2009astro2010S..82F, 2008PhRvD..78j3511P,2006ApJ...653..815M},
and possibly fundamental physics
\cite{2009astro2010S.234P,2006PhR...433..181F}.  There are currently
two standard approaches to the detection of the cosmological 21-cm
signal: measuring the redshift evolution of its sky-averaged value
(commonly labeled the ``global 21-cm signal'') \cite{2010PhRvD..82b3006P}, 
and measuring the statistical properties of the full three-dimensional
fluctuations in the 21-cm intensity \cite{2008PhRvD..78j3511P}.
The leading global experiment, EDGES, uses a single dipole to
constrain the duration of reionization \cite{2009arXiv0910.3010M, 2008ApJ...676....1B, 1999A&A...345..380S, 2010Natur.468..796B}.  
More advanced single-dipole experiments, such as a lunar satellite,
are currently being planned \cite{2011AAS...21710709B}.  Multi-dipole
interferometers such as LOFAR
\footnote{http://www.lofar.org}, MWA
\footnote{http://www.MWAtelescope.org}, GMRT
\cite{2010arXiv1006.1351P}, and PAPER \cite{2010AJ....139.1468P}, will
search for the fluctuations caused by ionized regions during the epoch
of reionization (EoR) and their 3D power spectrum
\cite{2006ApJ...653..815M, 2006ApJ...638...20B, 2004ApJ...608..622Z}.

In both approaches, a central technical challenge is to remove the
galactic synchrotron and other foregrounds from the cosmological 21-cm
signal (CS) associated with the reionization of hydrogen
\cite{2003MNRAS.346..871O}.  The mean value of the foregrounds are on
the order of hundreds of Kelvin even in quiet parts of the sky at
$\sim 100$MHz, while the magnitude of the CS is on the order of tens
of mK\cite{2006PhR...433..181F, 2006ApJ...653..815M, 2009arXiv0910.3010M}.  The foreground spectrum is smooth in frequency
relative to the CS, which makes foreground subtraction possible in
principle \cite{2007MNRAS.377.1043M, 2008ApJ...676....1B,2004ApJ...608..622Z, 2002ApJ...564..576D}.  Generically, the measured
signal can be fit by a slowly varying function in frequency (for
instance, a polynomial in frequency or log frequency
\cite{2009MNRAS.398..401L, 2008ApJ...676....1B}, or non-parametric
fits \cite{2009MNRAS.397.1138H}), with the residuals containing some
combination of fitting error and the fluctuating CS.  If the fit is
inadequate, residual foregrounds contaminate the CS.  If the fit too
closely matches the observed data, the CS will be removed together
with the foreground.  Several authors have already investigated the
optimal fitting techniques that offer a good compromise for the
tension between these opposing restrictions \cite{2008ApJ...676....1B, 2009MNRAS.397.1138H}.

So far, a single dipole antenna such as in EDGES constrained the
duration of reionization to be longer than $\Delta z > 0.07$
\cite{2010Natur.468..796B}. The need to fit the unknown foreground and
instrument response by a high-order polynomial in frequency suppresses
the ability to detect a gradual reionization process as a function of
redshift \cite{2008ApJ...676....1B, 2010PhRvD..82b3006P, 2010arXiv1005.4057P}.  Only a sharp change in the mean brightness
temperature as a function of frequency is detectable, making it
difficult to constrain the most likely theoretical scenarios
\cite{2009arXiv0910.3010M, 2010arXiv1005.4057P}.  Thus, at present, we
still do not have a 21-cm detection of the ``redshift of
reionization'' -- the point at which the universe was half ionized.

Given the above experimental status, we would like to examine whether a
modest interferometer (such as an early stage prototype of an
interferometer like MWA 32T) can do any better than a global
experiment like EDGES at detecting the redshift of reionization.
While an interferometer cannot measure the global signal, it could
trace evolution in the statistical properties of the 21-cm brightness
fluctuations with better calibration and foreground subtraction
prospects than for a single dipole experiment. In this paper, we
calculate the strength of the 21-cm signal in a given fiducial model
and give a plausibility argument that a modest (512 dipole)
interferometer with complete Fourier ($uv$) coverage should be
adequate to measure the redshift of reionization in that scenario.
Measuring $z_{\rm reion}$ would be a natural first target for 21-cm
surveys and an ideal way to demonstrate proof-of-concept for
low-frequency interferometers at an early stage.

We make several simplifications in our calculation.  Most importantly,
we assume full or nearly full $uv$ coverage, which eliminates the influence of sidelobes.  
Fortunately, this assumption
corresponds to the typical design goal of first generation
low-frequency arrays being built for EoR science
\cite{2008ApJ...680..962L}.  Under this assumption, our results do not
depend fundamentally on the details of tile placement and rotation
synthesis.  As long as any particular array is designed to have a
nearly complete $uv$ coverage, our calculation should be applicable.
It has also been shown in foreground and point source removal
studies that full $uv$ coverage greatly mitigates other instrumental
problems, which we also do not model explicitly.

Since we are interested in full $uv$-coverage, all of the information
collected by the interferometer is contained in the ``dirty map'' of
the sky at the maximum resolution of the instrument.  Since a
prototype instrument will not be able to probe down to the arcminute
scales of reionization (where the variations are the strongest) and
still maintain full $uv$ coverage, most of the information of interest
will be in the smallest scale available to us. The pixel-to-pixel
variance of a ``dirty map'' is a simple way to encapsulate the
information at the scale of interest \cite{2008MNRAS.389.1319J}.
While this does throw out some information, for instance the location
of the pixels and the longer-wavelength Fourier modes, it is a simple
statistic to work with.  If a detection can be shown to be feasible
with just the variance of the dirty map, then it will surely be
feasible with a more complete analysis.

Specifically, our approach is to calculate the standard deviation of the residuals
after foreground subtraction in each dirty map, as a function of
frequency.  What will remain in these residuals are the brightness
temperature fluctuations ($\sigma_{Tb}$) on top of a noisy spectrum and any fitting residuals.  We show that in
the regime where the spin temperature of the cosmic gas is much higher
than the CMB temperature ($T_s \gg T_{CMB}$), a peak in $\sigma_{Tb}$
is reached when the ionization fraction reaches $\sim 50\%$, as noted by others \cite{2008ApJ...680..962L},
providing a natural flag for $z_{\rm reion}$.  This holds when the
pixels of the map are independent, which is roughly valid as long as
their size is not less than the size of the ionized bubbles \cite{2006MNRAS.372L..43B}.

While $z_{\rm reion}$ has been constrained by the measurement of the
total optical depth for electron scattering with WMAP satellite, this
is in fact an integrated constraint on the entire history of
reionization \cite{2010arXiv1001.4538K, 2006PhR...433..181F}.  The
technique considered here picks out a specific redshift without
integrating along the line of sight.  By combining the two
measurements, it should be possible to constrain the duration of
reionization as well.

In order to estimate the prospects for this kind of measurement, we
consider 1D statistics representing the evolution of a pixel with
frequency (a beam).  We investigate how these single-beam
1D-statistics vary as a function of angular resolution (or ``beam
width'') and consider intermediate beam widths between the standard
global experiments and the full MWA or LOFAR arrays.  We also consider
the dependence on frequency resolution.  This is necessary because
both the thermal noise and brightness temperature fluctuations depend
strongly on the angular scales involved. Signal fluctuations increase
with higher resolution, because the instrument averages over less of
the sky.  However, this dependence is not trivial because the bubbles
have a finite size in real space and their locations are correlated.
Since the noise temperature varies according to the radiometer
equation, thermal noise increases rapidly at higher resolutions at a
fixed collecting area and integration time.  To calculate the expected
signal, we adopt a semi-numerical realization of reionization and
idealize 21-cm observations as pencil beams passing through the
simulation box.

The beam width dependence of $\sigma_{Tb}$ has been
previously studied with the aid of a radiative transfer simulation in
a cosmological box of $4h^{-1}$ comoving Mpc (cMpc) on a side
\cite{2004ApJ...608..611G}.  However, this size is an order of
magnitude too small to resolve the typical bubble sizes at the end of
reionization, and is thus of limited applicability.  Simulation boxes
of at least 100 comoving Mpc are necessary to reliably capture the
bubble size distribution at the end of reionization
\cite{2004ApJ...609..474B}.  We use a publicly available
semi-numerical code called {\it 21cm FAST} to overcome this hurdle
\cite{2007ApJ...669..663M, 2010arXiv1003.3878M}.

The outline of this paper is as follows.  In \S \ref{methodology} we
outline the methodology used as a function of beam width and frequency
resolution.  In \S \ref{singlebeam} we consider the signal in a
typical beam and calculate the effect of foreground subtraction on it.
In \S \ref{beamtobeam} we explore the statistics of an ensemble of
one-dimensional beams.  In \S \ref{subtract} we explore how to
optimize beam width and evaluate prospects for measuring $z_{\rm reion}$.  In \S \ref{instrument} we briefly consider instrumental issues and show that they should be surmountable in light of previous work.  Finally, \S \ref{concl} summarizes our main results and caveats.
\section{Methodology \label{methodology}}
In order to study the statistical properties of the CS, we simulate a
cosmological box using the publicly available {\it 21-cm FAST} code
\cite{2007ApJ...669..663M, 2010arXiv1003.3878M}.  This code generates
a random realization of the matter density field and then uses an
excursion-set approach to identify ionized regions.  It has tunable
cosmological parameters, ionization prescriptions, and several
optional features such as a halo-finder and a spin temperature
evolution tracker.  We do not use these optional features and instead simply use default cosmological and reionization
prescription settings, documented in Ref. \cite{2010arXiv1003.3878M}.
For our analysis, we use two different realizations, one with merely a linear density evolution
and the other with nonlinear evolution.  
Using both a linear and non-linear evolution enables us to resolve a
large volume and achieve high resolutions for our simulated beams in the
redshift range of $7<z<15$. 

Comparisons to radiative transfer simulations have shown that this
technique is reasonably accurate \cite{2007ApJ...669..663M, 2010arXiv1003.3878M}.  For our purposes, this method has the advantage
of being very computationally inexpensive while maintaining accuracy
at the tens of percent level \cite{2010arXiv1003.3455Z}, an advantage
shared by this class of programs \cite{2009MNRAS.393...32T}.

Our non-linear high resolution simulation involves a 300 cMpc
cosmological box with 600 pixels on a side, i.e.  0.5 cMpc per pixel
(corresponding to a frequency resolution of $\Delta \nu_{min} \approx 28$ kHz, or an angular resolution of $\theta_{min} \approx 0.2'$ in this redshift range).  Our larger, low resolution, linear realization contains 600 cMpc with a resolution of 200 pixels on a
side, or 3 cMpc per pixel (corresponding to $\Delta \nu_{min} \approx 170$kHz, or $\Delta \theta_{min} \approx 1.1'$).  We then generate
simulated sight-lines (``skewers'') of varying widths and
line-of-sight depths which start at a random location and
orientation. For simplicity, we adopt a square, top-hat
cross-sectional (transverse to the line-of-sight) shape to our
skewers and assume a number of beams for an instrument which has a 16 square degree field-of-view.

We then assume randomized periodic boundary conditions on the box, such
that if a skewer reaches the edge of the box, it re-enters the box at
a new random orientation, face, and location.  The redshift is
advanced in discrete time steps of $\Delta z = 0.25$ (corresponding to
a frequency interval of 3.5 MHz at the redshifts of interest), and the
sampling process resets after each time step.  We sample each redshift
in proportion to the number of real-space pixels which would be
crossed during that redshift interval. Until the redshift advances a
step, we linearly interpolate to calculate a frequency for that part
of the beam.

We assume that the beams have a fixed comoving width, each
corresponding to a fixed angular scale at the high redshifts of
interest (for $z \approx 10$, the variation in angular size for a
constant comoving width between $z=7$ and $z=12$ is $\lesssim 25\%$).
We adopt $\Lambda$CDM parameters of $\sigma_8 = 0.817$, $h=0.7$,
$\Omega_m = 0.28$, $\Omega_b = 0.046$, $\Omega_{\Lambda} = 0.72$,
consistent with the 7-year WMAP data \cite{2010arXiv1001.4538K}.
Throughout the paper, all length scales are in comoving units unless
otherwise specified.

\section{Single-beam properties \label{singlebeam}}

Each observing beam from an interferometer samples a new region of
space.  Therefore, one should expect cosmological fluctuations in the
differential brightness temperature within each beam in frequency
space.  The 21-cm line is observed in either emission or absorption,
depending on whether the local spin temperature, $T_s$, is above or
below the cosmic microwave background (CMB) temperature,
$T_{\gamma}=2.73{\rm K}\times (1+z)$.
The 21-cm fluctuations corresponding to peculiar
velocities, density fluctuations, and ionization fraction
fluctuations are described by the familiar equation, \cite{2006PhR...433..181F}
\begin{equation}
\begin{split}
\delta T_b \approx 9 x_{HI}(1+\delta)&(1+z)^{1/2}
\left(1-\frac{T_\gamma (z)}{T_s}\right) \\ & \times
\left(\frac{H(z)/(1+z)}{dv_{\parallel}/dr_{\parallel}}\right)~\mbox{mK} ,
\end{split}
\end{equation} 
where $\delta T_b$ is the differential brightness temperature,
$x_{HI}=1 - x_i$ is the neutral hydrogen fraction (by volume),
$\delta$ is the overdensity, and $H(z)$ is the Hubble parameter at
redshift z which is equal to the line-of-sight velocity gradient
$(1+z)dv_{\parallel}/dr_{\parallel}$ in the absence of peculiar
velocities \cite{2006PhR...433..181F}.  In the linear regime
applicable on $\gtrsim$Mpc scales, density fluctuations can be ignored
relative to the order unity fluctuations in the neutral fraction
$x_{HI}$.  If peculiar velocities are also ignored and $T_s\gg T_\gamma$, then 
\begin{equation}
\label{simpletb}
\delta T_b \approx 9 x_{HI}(1+z)^{1/2}~\mbox{mK.} 
\end{equation}
The brightness temperature fluctuations over scales of $\gtrsim$ Mpc
can be modeled as overlapping, growing spherical
``bubbles'' of ionized hydrogen which form around the first ionizing
sources
\cite{2006PhR...433..181F,2007ApJ...654...12Z,2007MNRAS.377.1043M, 2009arXiv0910.3010M, 2004ApJ...613...16F}.

The 21-cm signal from reionization gives each beam a complex
structure, because each ionized bubble can occupy part or all of the
beam both along the line of sight and in the plane of the sky.  In addition,
overlapping bubbles might correlate $\delta T_b$ among adjacent beams
which belong to the same bubble \cite{2004ApJ...608..622Z}.  This
correlation can be characterized by the form
\cite{2004ApJ...613...16F}:
\begin{equation}
\label{correlationlength}
\langle x_i(r) x_i(r') \rangle = \bar{x}_i^2+(\bar{x}_i - \bar{x}_i^2)f(r/R_c),
\end{equation} 
where $r \equiv |r-r'|$ and $f(r/R_c)$ parametrizes the effect of the
finite characteristic size $R_c$ of the bubbles on the correlation.  The characteristic
scale of the H II bubbles has been shown to depend primarily on the mean
ionization fraction and, to a lesser degree, on other parameters
\cite{2007MNRAS.377.1043M}.  Equation \ref{correlationlength} implies
that if $r\gg R_c$, $f(r/R_c)$ approaches zero and the variance of 
$\delta T_b$ is maximized when the ionized fraction is 50\%
(ignoring density fluctuations).  We will use this feature as our
marker for the reionization redshift, $z_{\rm reion}$, and verify how valid this is
 at the relevant scales using {\it 21-cm FAST}. 

\begin{figure}[h!]
\includegraphics[width=3.25in]{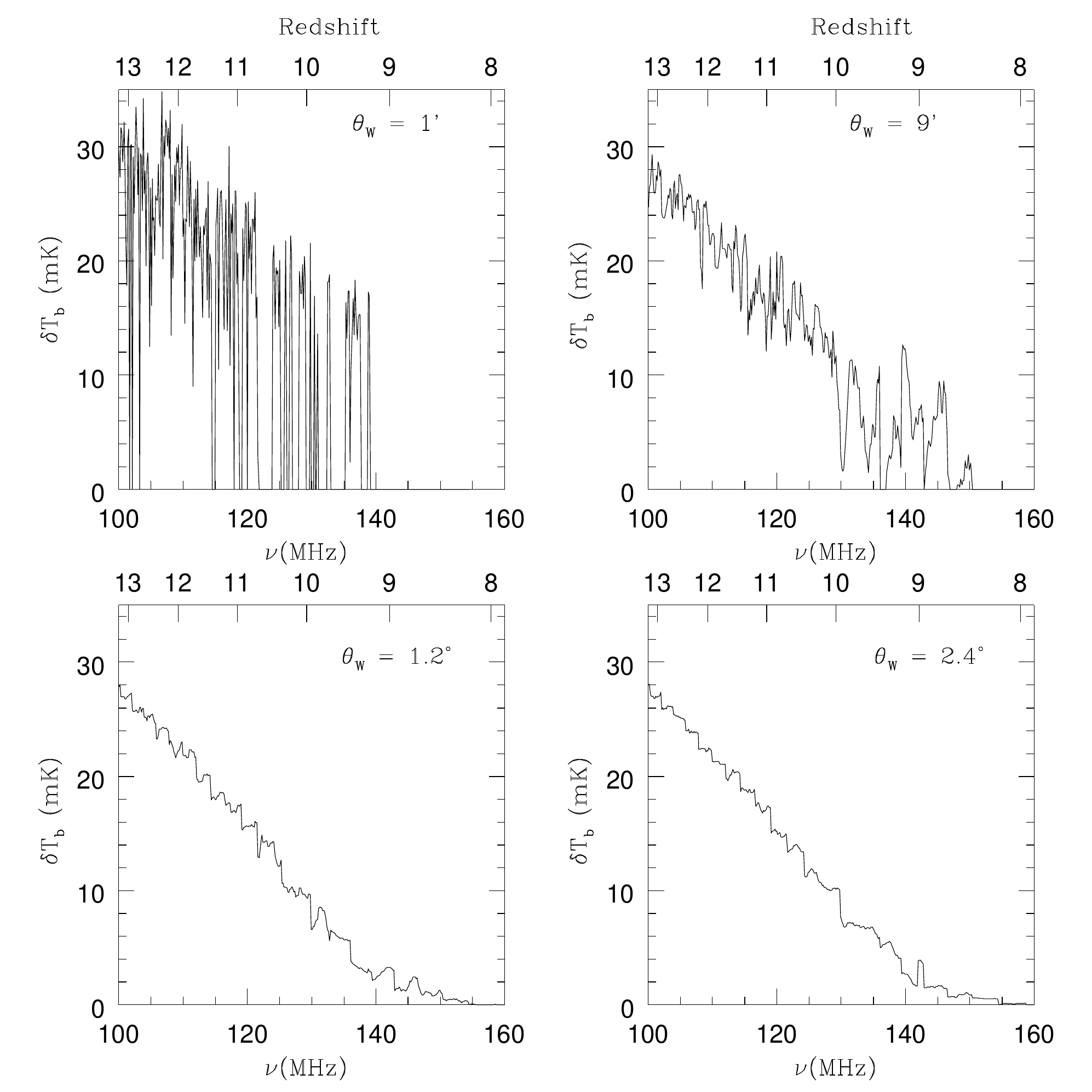}
\caption{\label{singlebeamsnake} Typical skewers with beam widths
$\theta_w$ of 1' (top left), 9' (top right), $1.2^{\circ}$ (bottom
left), and $2.4^{\circ}$ (bottom right), with $\Delta \nu \approx 30 \rm{kHz}$.  Increased beam width reduces the average level at which
the signal varies by averaging over a bigger area of sky at each
frequency interval.  There is no thermal or instrumental noise in
these examples. As others have found, foreground subtraction on these
signals with high order polynomials does not remove much of the power
from the line-of-sight fluctuations.}
\end{figure}

Figure \ref{singlebeamsnake} shows typical skewers (beams) with four
different widths.  Two effects are evident from these examples.
First, the underlying bubble structure causes the CS to vary with
frequency much more rapidly than its mean behavior, even
though reionization itself is progressing slowly.  Secondly, it is visually
obvious that while the mean behavior of these signals is the same,
the magnitude of the variations from the mean depends strongly
on beam width, which we consider more carefully in \S
\ref{beamtobeam}.  

\begin{figure}[h!]
\includegraphics[width=3.250in]{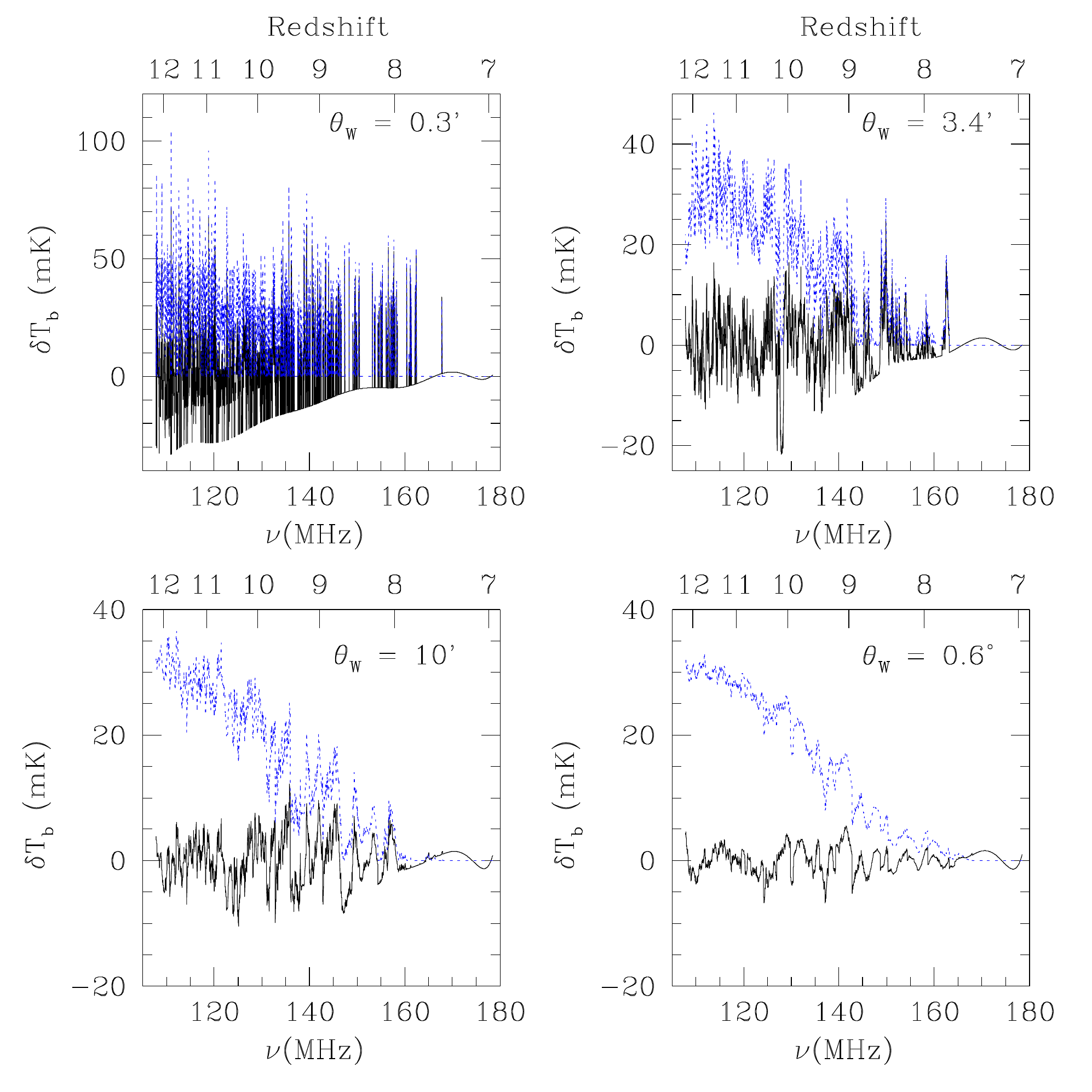}	
\caption{\label{foregroundsubtract} Signal after subtracting out a 12
order polynomial fit to the spectral data as a function of beam width.
In each panel, the dashed top curve shows the original signal without
any foregrounds, while the bottom solid curve shows the residuals
after foreground subtraction.  The beam widths at $z=9$ are 0.3' (top
left), 3.4' (top right), 10' (bottom left), and $0.6^{\circ} $ (bottom
right).  Note the large difference in the residual signal amplitude
between the different panels.}
\label{Fig2}
\end{figure}

In order to recover these skewers, it is necessary to subtract out the foregrounds using some kind of fit or model of them.  
While others have done this in much greater detail, we briefly verify that this is possible in principle in our analysis - i.e. even a high-order polynomial will not remove all the power from the CS fluctuations in frequency space.  In order to see what part of the signal would remain
after a foreground subtraction, one can add an approximation of the
noise, and then fit the combined signal and subtract out the fit.  We
approximate the signal in quiet portions of the high-latitude radio sky
as a power law in frequency \cite{2006PhR...433..181F}:
\begin{equation}
\label{skytemp}
T_{sys} \approx T_{sky} \approx \mbox{180} \left( \frac{\nu}{\mbox{180
MHz}} \right)^{-2.6}~\mbox{K} 
\end{equation}
After adding this mean behavior to our individual beams, we 
subtract out a 12th order polynomial in frequency to fit the
signal, and plot the residuals in Figure \ref{foregroundsubtract}.


It is evident from Figure \ref{Fig2} that the fitting residuals do not in principle dominate the signal if all we are concerned about is
the magnitude of the sky temperature varying with frequency.  We did not make a fully adequate foreground model here, but we find fitting residuals to be at about $\approx 0.75$ mK for the most relevant angular resolution of $1.2^{\circ}$ with a 3rd order polynomial, or $\approx 0.4$ mK with a 3rd and 12th order polynomial, with foreground subtraction performed before binning of adjacent frequency bins.

Of more concern is whether foreground subtraction is still effective after including point sources and other frequency dependent
distortions.  This question is studied much more carefully by other
authors \cite{2009ApJ...695..183B, 2009MNRAS.394.1575L} in the context
of a full first generation array.  With 3rd order polynomials, Bowman et al. find a best case scenario of 1 mK residuals after foreground subtraction.  The magnitude of these residuals sets a systematic bound on the minimum signal strength that will be distinguishable from foreground subtraction errors (as discussed in \S \ref{subtract}).  We take the approach of trying to keep the magnitude of the signal above 1 mK while noting that increasing the order of the polynomial could possibly mitigate this problem (with caveats about losing more of the long-wavelength signal).

It remains to be seen whether a
prototype array will be able to overcome these issues, although we are
hopeful that it will (see \S \ref{instrument}). At any rate, our
results demonstrate that very little power from the CS (from the scales of interest to us) is removed
within a single beam, even by a high-order polynomial at degree-scale
resolution.  Figure \ref{Fig2} provides a simple way to understand why
global experiments have a difficult time detecting reionization in a
gradual reionization case: as the resolution scale increases, the
fluctuations get smaller, making it more difficult to separate the
post-subtraction fluctuations from noise alone.

Thus, our post-subtraction signal is composed of these
fluctuations plus thermal noise in the antenna which also generates
deviations. To demonstrate that we can measure these fluctuations
accurately, we need to quantify the thermal noise in the antenna due
to the mean sky temperature and the magnitude of deviations as a
function of beam width and frequency resolution.  We address the
latter issue first.

\section{Beam-to-beam variance \label{beamtobeam}}

Next, we consider an ensemble of beams that are sufficiently narrow to
probe fluctuations caused by inhomogeneous reionization.  We can then
address the ``beam-to-beam'' variance ($\sigma^2_{Tb}$), namely the
variance in the signal at a given frequency bin over an ensemble of
beams.  This is a natural quantity to consider for an interferometer
which observes a number of beams, $N_{\rm beams} \approx \Delta\Omega / \theta_w^2$, where $\Delta\Omega$ is the solid angle of the field of
view and $\theta_w$ is the width of each square beam.  The
beam-to-beam variance or standard deviation is useful because
interferometers only measure the deviations from the mean signal.

\begin{figure}[h!]
\includegraphics[width=3.25in]{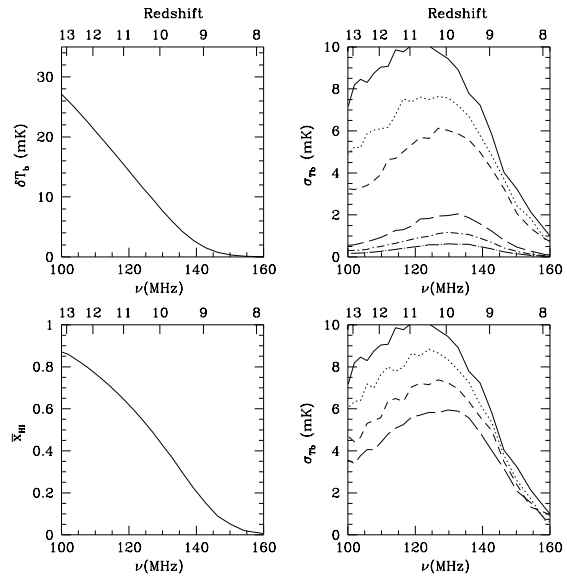}
\caption{\label{snake4up} {\bf Top left}: Evolution of the mean
$\delta T_b$ in a 600 cMpc box. {\bf Top right:} The standard
deviation of $\delta T_b$ as a function of redshift for 6 beam widths
(top to bottom: 1', 2', 4', $0.6^{\circ}, 1^{\circ}, 2^{\circ}$).  The
frequency resolution was $\sim 150 $ kHz in all cases. Note that for
larger beam widths, the maximum variance occurs at a
redshift of about 10 when the ionization fraction is 0.5. {\bf Bottom
left}: Evolution of the mean neutral fraction in a 600 cMpc box. {\bf Bottom right:} The standard deviation of $\delta T_b$ as a function of
redshift for 4 different frequency resolutions $\Delta \nu$ (top to
bottom: 150 kHz, 300 kHz, 600 kHz, 1.2 MHz).  The beam width was 1' in
all cases.  Again, the peak occurs at $z\sim 10$ for large $\Delta \nu$.}
\end{figure}

Since the number of regions averaged over is a function of the
resolution, it is important to consider the beam width and frequency
dependence of the beam-to-beam variance.  If there were no
correlations among nearby regions (i.e., bubbles were much smaller
than the regions under consideration, and regions were not
correlated), the sample standard deviation would fall off as $1/\sqrt{N_{\rm pixels}}$ where $N_{\rm pixels}$ is the number of pixels.
However, inside of large ionized regions, there is near-perfect
correlation among nearby pixels. Moreover, because the ionized regions
themselves are correlated, the maximum variance may occur in an
intermediate regime, where just a few bubbles are being averaged
over. The largest ionized bubbles obtain a size of $\sim 100$ cMpc at
the end of reionization \cite{2004Natur.432..194W}.  This corresponds
to $\sim 0.6$ degrees at z=9.  Thus, the exact dependence of
$\sigma_{Tb}$ on resolution is both non-trivial and important to this
measurement.  It is most convenient to use {\it 21-cm FAST} to
estimate this standard deviation of $\delta T_b$ as a function of both
frequency and angular scale.  An illustration of this dependence is
depicted in Figure \ref{snake4up}.

As expected, the beam-to-beam standard deviation decreases more
slowly than the $1/ \sqrt{N_{\rm pixels}}$ statistics would
predict.  For instance, in the bottom right panel of
Figure \ref{snake4up}, it takes a factor of $\sim 8$ more pixels to
reduce the signal by a factor of $\sim 0.5$.  Each pixel is being
randomly chosen, but clusters of pixels chosen together will be
correlated because of the finite size of the ionized bubbles.  As shown in
the next section, this significantly boosts our
signal-to-noise ratio, because the experiment requires seeing the peak
of this signal against fluctuations which do obey $1/ \sqrt{N_{\rm pixels}}$ statistics.

The location of the peaks on the right hand side of Figure
\ref{snake4up} does seem to be moderately scale-dependent.  The large
component of the variance which originates from uncorrelated bubbles
peaks at a neutral fraction of 50\%.  However, on any scale, there is an
additional component originating from the second term of equation
(\ref{correlationlength}), which determines the redshift at which the
variance peaks.  The inferred redshift of reionization (defined by the
neutral fraction being 50\%) has therefore some model-dependent
uncertainty. For any specific model of reionization, this
uncertainty can be removed by a comparison to the model's version of
Figure \ref{snake4up}.

\section{Measuring the signal against thermal noise \label{subtract}}

As mentioned above, our approach is to measure the standard deviation
of each dirty map as a function of frequency.  The standard deviation
of the ensemble of observed beams includes a few different components:
the thermal noise, the standard deviation of $\delta T_b$ which peaks
at some redshift, and any fitting errors, which we ignore.  The
variances of noise and signal at any redshift add directly by the
Bienaym\'{e} formula (since the noise is assumed to be uncorrelated with the
signal):
\begin{equation}
\label{quadrature}
\sigma^2_{\rm dirty}(\nu) = \sigma^2_{Tb}(\nu)+\Delta T_{meas}^2(\nu) + {\rm systematic}
\end{equation}
The standard deviations of $\delta T_b$
($\sigma_{Tb}$) were calculated in \S \ref{beamtobeam} and are
expected to peak near $x_i=0.5$, providing a flag of $z_{\rm reion}$.
This peak sits on top of the noise spectrum, which should follow
Eqs. (\ref{quadrature}) and (\ref{thermalradiometer}).  In an
interferometer imaging experiment, the noise is obtained from the
radiometer equation.  Using typical values for the experimental
parameters of 21-cm observations, the noise temperature is given by,
\cite{2006PhR...433..181F}:
\begin{equation}
\label{thermalradiometer}
\begin{split}
\Delta T^N \approx 2 \rm{mK} \left(\frac{A_{tot}}{10^5
m^2}\right)^{-1} \left(\frac{10'}{\theta_w}\right)^{2} \\
\left(\frac{1+z}{10}\right)^{4.6} \left(\frac{1 \rm{MHz}}{\Delta \nu}
\frac{100 \rm{hr}}{t_{int}}\right)^{1/2} ,
\end{split}
\end{equation}
where $A_{\rm tot}$ is the effective collecting area of the
interferometric array, $\theta_w$ is the angular resolution, $\Delta \nu$ is the bandwidth, and $t_{int}$ is the integration time.  At a
fixed collecting area and integration time, the noise
temperature is very sensitive to the angular resolution and $\Delta \nu$. 
In order to extract the standard deviation excess caused by the 21-cm fluctuations from $\sigma_{\rm dirty}(\nu)$, one can subtract off a model or power law fit to the $(\Delta T^N)(\nu)$ spectrum.  
\begin{equation}
\sigma^2_{\rm Tb , meas}(\nu) = \sigma_{\rm dirty}(\nu)^2-\Delta
T_{\rm fit}^2(\nu).
\end{equation}  
We will assume that this fit will only leave residuals due to the
finite sampling errors of measuring a variance.  These sampling errors
are the ``effective noise'' $\Delta T_{\rm eff}$, which might cause us
to make a mistake in determing the true variance at any frequency bin.
For a Gaussian probability distribution in the $N_{\rm beams} \gg 1$
limit, the average deviation between the measured noise standard
deviation and the true variance of the noise is given by a standard
statistics result \cite{VarianceBook}:
\begin{equation}
\sqrt{\langle var(s_{\Delta T})\rangle} = \frac{1}{\sqrt{2 N_{\rm beams}}} (\Delta T^N)
\end{equation} 
Similarly, the sample variance of the signal would be:
\begin{equation}
\sqrt{\langle var(\sigma_{Tb})\rangle} = \frac{1}{\sqrt{2 N_{\rm beams}}} (\sigma_{Tb})
\end{equation} 
Therefore, we can define the effective noise as
\begin{equation}
\Delta T_{\rm eff} = \sqrt{\frac{1}{2 N_{\rm beams}}((\Delta T^N)^2+\sigma_{Tb}^2)}
\end{equation} 
Under these assumptions, we define the
effective signal-to-noise ratio as
\begin{equation}
S/N_{\rm eff} = \sqrt{\frac{\sigma^2_{Tb}}{\Delta T_{\rm eff}^2}} = \frac{\sigma_{Tb}}{\sqrt{((\Delta T^N)^2+\sigma_{Tb}^2)/2}}.
\end{equation}

\begin{figure}[h!]
\includegraphics[width=3.25in]{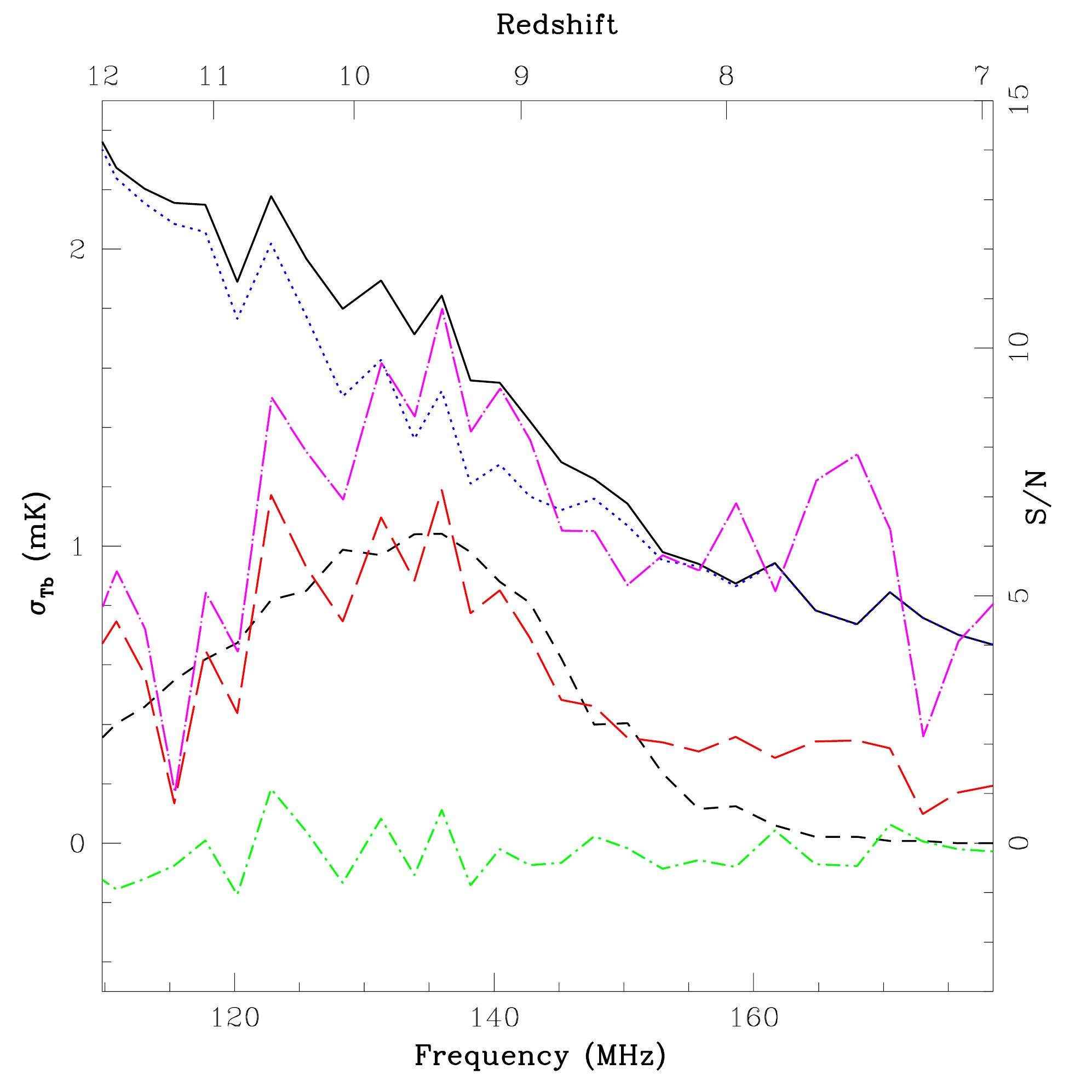}
\caption{\label{examplevariancedata} An illustration of the different
components in the measurement of the reionization redshift $z_{\rm reion}$, for the modest array parameters: $A_{tot} = 500 ~ \rm{m}^2$
(roughly 32$\times$16 dipoles), $\theta_w=1.2^{\circ}, \Delta \nu = 2.4 \rm{MHz}$, and $t_{int}=\rm{500~ hours}$.  The beam-to-beam {\it rms} variation (solid black top curve) is composed of both the variance due to thermal noise (dotted blue curve) and the signal caused by the bubbles (black short-dashed curve).  If the expected variance from noise is subtracted off from the measured variance, all that remains are the signal and sampling errors as fitting residuals of the noise profile (red long-dashed curve).  If there were no signal at all, the expected profile final result would only be due to sampling errors (bottom green dot-dashed curve).  The signal-to-noise (purple dot-dashed curve) is calculated as defined in the text at each frequency bin.  These noise estimates assume a ``quiet'' high latitude portion of the sky described by equation \ref{skytemp}.}
\label{Fig4}
\end{figure}

Our procedure is illustrated in Figure \ref{examplevariancedata}.  We
infer the redshift of reionization from the highest
signal-to-effective-noise point on the curve, which, as mentioned
before, is a somewhat model-dependent procedure that could be improved
given some model for reionization.  However, it assumes only the
existence of a peak in the $T_s \gg T_k$ regime.  In practice, 
 systematics are likely to be the actual limiting factor in the
choice of the resolution.

We note that a higher angular resolution (small beam width $\theta_w$)
increases the number of beams and the absolute magnitude of the
signal.  However, the signal-to-noise ratio increases at larger values
of $\theta_w$ and $\Delta \nu$ up to a point, because the effective
noise goes down more rapidly than the signal standard deviation up to
the characteristic size of an individual ionized bubble.  Furthermore,
the scale must be big enough for the beams to be at least roughly
independent, or the peak will not occur when the neutral fraction has
reached 50\%.  Thus, to measure $z_{\rm reion}$, it is ideal to probe
degree scales, at which signal-to-effective-noise-ratio is high, the beams
are roughly independent, and the signal is comparable to the absolute
magnitude of the noise and larger than systematic effects.  In
practice, we find this to be between $\sim 0.6^{\circ}$ and $\sim 1.2^{\circ}$ at $\Delta \nu=2.4$ MHz where the beams are roughly
independent, $\sigma_{Tb}$ is between 3 mK and 1 mK, and the signal to
noise at the peak is between 11 and 13.  Systematics will most likely
limit us because foreground subtraction with low order polynomials
most likely cannot be accomplished at the better than 1 mK level
\cite{2009ApJ...695..183B}.  If a stronger absolute magnitude of the
signal is required, smaller angular resolutions can be used and
compensated for by longer integration times.

To estimate the error in $z_{\rm reion}$, we randomly draw Gaussian noise, add it to the sample realizations shown in Figure
\ref{examplevariancedata}, and calculate the standard error in the
location of the highest peak after 100 Monte Carlo trials of this procedure.  Our result
is that for $t_{int}=500$ hours, $A_{tot} = 500 ~ \rm{m}^2 $
(corresponding to MWA with 32 tiles of 16 dipoles each),
$\theta_w=1.2^{\circ}$, and $\Delta \nu = 2.4$ MHz, one can achieve a
system noise temperature of $\Delta T^N = 0.94$ mK and $\Delta T^N_{\rm eff}=0.12 {\rm mK}$ at $z_{\rm reion}$.  With these
parameters, one can keep the CS above 1 mK and make a $4 \sigma$
detection of $z_{\rm reion}$ with a standard error of $\Delta z_{\rm reion} = 0.25$.  Depending on realistic integration time constraints
and systematic issues caused by antenna layout, optimal parameters may
vary for actual experiments. The above numbers indicate that such a
measurement is possible with current technology.  However, these numbers
ignore systematics from fitting errors, any residual discrepancy
between the peak and the redshift of reionization, and RFI.

\begin{figure}[h!]
\includegraphics[width=3.25in]{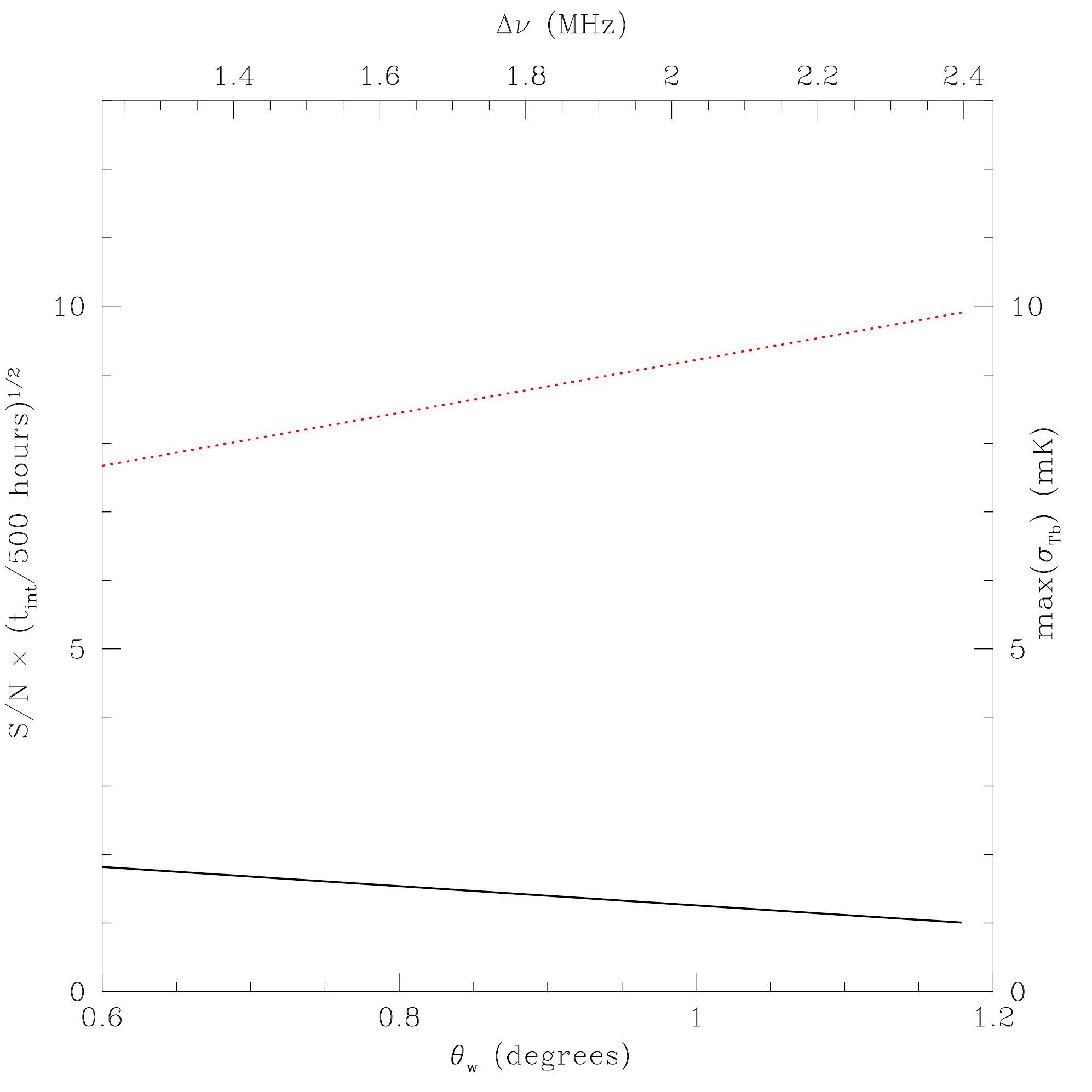}
\caption{\label{newoptimal} Estimated signal-to-noise ratio (dotted
line) and absolute magnitude of signal (solid line) vs. beam width and
band width. The angular width is increased along with frequency width
to keep the ratio of the real space size of the pixels the same in
different directions.  The expected signal-to-noise ratio that can be
achieved is set by the systematics of the experiment (left axis), as
the absolute magnitude of $\sigma_{Tb}$ needs to stay above the
instrument systematic uncertainties (involving, e.g., residuals from
foreground subtraction (right axis).  In accordance with the
radiometer equation, we assume that these limits will allow a signal
of greater than 1 mK to be detectable.  We do not show values of beam
width for which we do not expect the beams to be independent, because this 
may introduce a non-negligible sampling error in $\sigma_{Tb}$.}	
\label{Fig5}
\end{figure}

\begin{figure}[h!]
\includegraphics[width=3.25in]{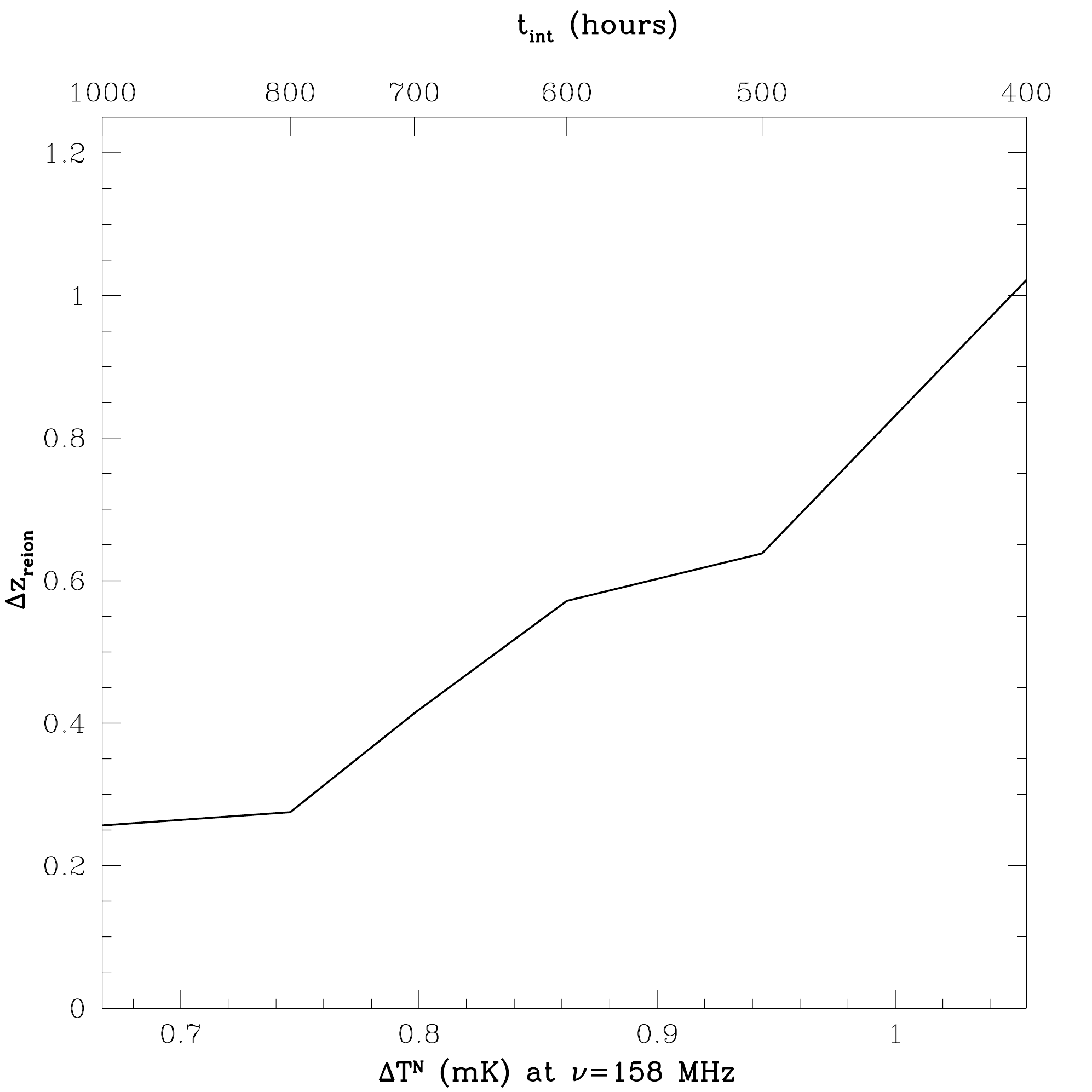}
\caption{\label{deltazvsnoise} The standard error in the global
redshift of reionization, $\Delta z_{\rm reion}$, as a function of
total noise temperature at an observing frequency of 158 MHz, with
$\theta_w = 1.2^{\circ}$ and $\Delta \nu = 2.4$ MHz for $A_{tot} \approx 500 {\rm m}^2$.  The integration time in hours is given on the
top axis.  $\Delta z_{\rm reion}$ is calculated by taking the standard
deviation of fifty Monte Carlo trials, following the procedure
illustrated in Figure \ref{Fig4}.  These errors ignore the systematic considerations discussed in the text.}
\label{Fig6}
\end{figure}

A plot of the signal-to-noise ratio and absolute amplitude of the
signal, $\max(\sigma_{Tb})$, versus increasing beam width and $\Delta \nu$ is shown in Figure \ref{newoptimal}.  In Figure
\ref{deltazvsnoise}, we show how strong the constraint is on $\Delta z_{\rm reion}$ as a function of integration time for the observing
parameters we chose.  

\section{Practical Considerations \label{instrument}}

So far, our estimates have been restricted to mostly theoretical considerations about signal and thermal noise.  In practice, however, detailed foreground and instrumental considerations are critical to determining the plausibility of such a measurement of $z_{\rm reion}$.

The chief analytical technique which will reduce these effects is the standard signal fitting and subtraction technique.  Each pixel can be independently fitted with a suitable function (polynomial or otherwise) as a function of frequency, and contributions from unresolved point sources, galactic synchrotron emission, and frequency-dependent instrumental response should be reduced, leaving only the fast-varying CS \cite{2008ApJ...676....1B, 2007MNRAS.377.1043M, 2008ApJ...676....1B, 2004ApJ...608..622Z, 2002ApJ...564..576D, 2009MNRAS.398..401L}.

However, it is also conceivable that a frequency-dependent point
spread function (PSF) or field-of-view (FOV) could mix power from
nearby pixels in a frequency-dependent way, and create artificial
peaks in a measurement of $\sigma_{Tb}$ versus frequency.  This is of
particular concern if the $uv$ plane is sparsely filled, which would
make the PSF deviate strongly from a $\delta$ function.  Also, if the
$uv$ plane is unfilled, pixels which may not be physically correlated
due to the bubbles may become correlated due to under sampling. It is
impossible to have more independent pixels than there are independent
baselines, whatever the nominal angular resolution.

Fortunately, we can rely on past work to show that this is unlikely to
be an issue even for smaller prototype experiments.  Bowman et al. found that, for a full MWA, a third order
polynomial does an excellent job of subtracting foreground
contamination for baselines where the $uv$ coverage was complete \cite{2009ApJ...695..183B}.
Bowman et al. also found that for these larger angular scales, with
uniform weighting of visibilities, the PSF was effectively a clean
$\delta$ function with no sidelobe confusion.  Similar results were
found by Liu et al. \cite{2009MNRAS.394.1575L}.

In particular, using the formalism from Bowman et al.
\cite{2009ApJ...695..183B}, we can estimate the variance due to
sidelobes in the synthesized array beam
\begin{equation}
\sigma^2_D \sim \sigma^2_S (1 + B^2_{rms} \Omega_P / \Omega_B)
\end{equation}  
where $B_{rms}$ is the rms value of the beam response relative to the
peak, $\Omega_P$ corresponds to the FOV and $\Omega_B \sim \theta_D^2$
is the solid angle of a resolution element.  For MWA 32T, the value of
$B_{rms}$ is $\sim 1\%$, and $ B^2_{rms} \Omega_P / \Omega_B \sim 10\%$ \cite{JuddPersonal}.  This implies the sidelobe confusion will
be 10\% of the sky variance, so sidelobe rms will be 5\% of the sky
rms, which is about 10K.  In other words, the sidelobe rms will be
$\sim 500$mK, which is a reasonable worst-case scenario, and in
practice one should be able to do better once foreground subtraction
is taken into account.

\begin{figure}[h!]
\includegraphics[width=3.25in]{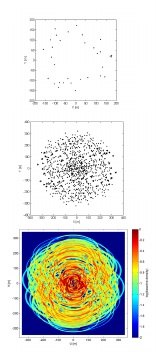}
\caption{\label{uvcoverage} Simulated $uv$ coverage after rotation
synthesis for MWA 32T. The upper panel shows the configuration
(in $x$-$y$ coordinates) of MWA 32T tiles, and the middle panel shows a snapshot of the corresponding baselines.  The
lower panel illustrates the tracks of the dipoles (due to Earth
rotation) in the Fourier $u$-$v$ coordinates, and illustrate that
$\sim 60\%$ of the $uv$ plane is filled after rotation synthesis with this prototype.  The details of the attenna layout are not used in the present calculation.
Plots are courtesy of J. Bowman \cite{JuddPersonal}.}
\label{Fig7}
\end{figure}

Therefore, one could hope an instrument with good $uv$ coverage at large angular
scales should be able to defeat these instrumental concerns.  This is not a merely hypothetical scenario, as the $uv$
plane should be reasonably well filled (about 60\%) for a reference
configuration of MWA 32T \cite{JuddPersonal}, as illustrated in Figure
\ref{Fig7}.  Thus, it seems plausible that MWA 32T itself may have a
good chance of detecting this signal, and an instrument with even better UV coverage would be more ideal.  Perhaps LOFAR $uv$ coverage
\cite{2009MNRAS.394.1575L} could be tuned to a similar performance
level if long baselines were masked from the analysis to keep the $uv$ plane filled.

Since this approach has not yet been tested in practice there could be
unknown instrumental issues, such as calibration issues for 32T, a
non-flat bandpass (which would require a higher order polynomial), or
issues with the beam response pattern \cite{JuddPersonal}.  There is
certainly room for more study of these issues from an instrumental
point-of-view.  But it seems reasonable to hope that dense $uv$
coverage of a compact core will ensure that frequency-dependent
instrumental response will not hinder polynomial foreground
subtraction.

\section{Conclusions}
\label{concl}

We have demonstrated in Figures \ref{Fig4}, \ref{Fig5} and \ref{Fig6}
that a modest interferometer with an optimal beam width and frequency
resolution can statistically detect the global redshift of
reionization, $z_{\rm reion}$.  The standard deviation of dirty
maps as a function of frequency at $1^{\circ}$ scales shows that detection prospects are plausible and that
a conventional power-spectrum analysis could probably do even better. In a single-dipole
experiment, current foreground removal strategies for global 21-cm
signal remove much of the global signal itself unless reionization
occurred over an unusually narrow redshift interval.  
Although the global signal has many interesting properties other than
just the redshift of reionization, it can only be used to rule out
fast reionization signal with the EDGES experiment (given current
calibration).  First-generation interferometers capable of measuring
the 21-cm power spectrum in all its detail may still be several years away, but a detection may be possible now.

We find that prototypes of interferometers designed to measure the
power spectrum should be able to measure the global redshift of
reionization without a large investment of time or funds, based on the
technique outlined in this paper.  We quantified the observational
prospects in a particular reionization scenario and showed that
intermediate beam widths of $\sim 1.2^\circ$, integrated over hundreds
of hours at MHz resolutions and $\sim 500$ $\rm{m}^2$ of effective
area should allow for a statistical detection of the global redshift
of reionization if instrumental and systematic challenges are
overcome.

It is difficult to predict how instrumental challenges, such as
removal of point sources, uneven instrumental response, RFI, and other
issues would play out for the measurement described here.  But work done by others
suggests that dense $uv$ coverage with a compact core will mitigate many potential issues, 
and that foreground subtraction should work at least as
effectively as it does for an EDGES-like global experiment.  

Although the power spectrum of 21-cm fluctuations at all scales provides a more
detailed probe of reionization, the 1D flag of $z_{\rm reion}$
considered here represents a ``zeroth order'' way to see that even prototype scale instruments can constrain reionization.  In principle, one might be able to use the
width of the $\sigma_{Tb}$ peak to constrain the duration of
reionization.  This would be a simple way to demonstrate the feasibility of the interferometric approach to the
emerging frontier of 21-cm cosmology.  \\

\section*{Acknowledgments}

This work was supported in part by NSF grants AST-0907890, and NASA
grants NNX08AL43G and NNA09DB30A.  We thank Gianni Bernardi, Judd Bowman, Max Lavrentovich, Matt Mcquinn, David Mitchell, and Jonathan Pritchard for useful discussions and comments.  We would like to especially thank Judd Bowman for providing guidance on instrumental issues and providing MWA 32T information, including Figure 7.

\def\aj{AJ}                   
\def\araa{ARA\&A}             
\def\apj{ApJ}                 
\def\apjl{ApJ}                
\def\apjs{ApJS}               
\def\ao{Appl.~Opt.}           
\def\apss{Ap\&SS}             
\def\aap{A\&A}                
\def\aapr{A\&A~Rev.}          
\def\aaps{A\&AS}              
\def\azh{AZh}                 
\def\baas{BAAS}               
\def\jrasc{JRASC}             
\def\memras{MmRAS}            
\def\mnras{MNRAS}             
\def\pra{Phys.~Rev.~A}        
\def\prb{Phys.~Rev.~B}        
\def\prc{Phys.~Rev.~C}        
\def\prd{Phys.~Rev.~D}        
\def\pre{Phys.~Rev.~E}        
\def\prl{Phys.~Rev.~Lett.}    
\def\pasp{PASP}               
\def\pasj{PASJ}               
\def\qjras{QJRAS}             
\def\skytel{S\&T}             
\def\solphys{Sol.~Phys.}      
\def\sovast{Soviet~Ast.}      
\def\ssr{Space~Sci.~Rev.}     
\def\zap{ZAp}                 
\def\nat{Nature}              
\def\iaucirc{IAU~Circ.}       
\def\aplett{Astrophys.~Lett.} 
\def\apspr{Astrophys.~Space~Phys.~Res.}
\def\bain{Bull.~Astron.~Inst.~Netherlands} 
\def\fcp{Fund.~Cosmic~Phys.}  
\def\gca{Geochim.~Cosmochim.~Acta}   
\def\grl{Geophys.~Res.~Lett.} 
\def\jcp{J.~Chem.~Phys.}      
\def\jgr{J.~Geophys.~Res.}    
\def\jqsrt{J.~Quant.~Spec.~Radiat.~Transf.}
\def\memsai{Mem.~Soc.~Astron.~Italiana}
\def\nphysa{Nucl.~Phys.~A}   
\def\physrep{Phys.~Rep.}   
\def\physscr{Phys.~Scr}   
\def\planss{Planet.~Space~Sci.}   
\def\procspie{Proc.~SPIE}   

\let\astap=\aap
\let\apjlett=\apjl
\let\apjsupp=\apjs
\let\applopt=\ao

\let\astap=\aap
\let\apjlett=\apjl
\let\apjsupp=\apjs
\let\applopt=\ao

\bibliography{21cmbubble7}

\end{document}